\documentclass[11pt]{article}
\usepackage{graphicx}
\usepackage{amsmath}
\usepackage{amssymb}
\usepackage{amsxtra}

\usepackage{amsfonts}
\usepackage{slashed}

\begin{document}

\begin{center}
EXTENDING STAROBINSKY INFLATIONARY MODEL \\ IN GRAVITY AND SUPERGRAVITY
\vglue.2in

 S.V. Ketov \\ 
 Research School of High Energy Physics, Tomsk Polytechnic University,  Lenin ave. 30, Tomsk 634050, Russia \\ Physics Department, Tokyo Metropolitan University, Minami-ohsawa 1-1, Hachioji-shi, Tokyo 192-0397, Japan \\ 
 Kavli Institute for the Physics and Mathematics of the Universe (IPMU), The University of Tokyo, Chiba 277-8568, Japan \\
 ketov@tmu.ac.jp  \\
 
 \vglue.2in
 
  M.Yu. Khlopov\\ 
  Institute of Physics,
Southern Federal University\\
Stachki 194,
Rostov on Don 344090, Russia\\
khlopov2apc.in2p3.fr

\end{center}

\vglue.2in

\begin{abstract}
We review some recent trends in the inflationary model building, the supersymmetry (SUSY) breaking, the gravitino Dark Matter (DM) and the Primordial Black Holes (PBHs) production in supergravity. The Starobinsky inflation can be embedded into supergravity when the inflaton belongs to the massive vector multiplet associated with a (spontaneously broken) $U(1)$ gauge symmetry. The SUSY and R-symmetry can be also spontaneously broken after inflation by the (standard) Polonyi mechanism. Polonyi particles and gravitinos are super heavy and can be copiously produced during inflation via the Schwinger mechanism sourced by the Universe expansion. The overproduction and instability problems can be avoided, and the positive cosmological constant (dark energy) can also be introduced.  The observed abundance of the Cold Dark Matter (CDM) composed of gravitinos can be  achieved in our supergravity model too, thus providing the unifying framework for inflation, supersymmetry breaking, dark energy and dark matter genesis. Our supergravity approach may also lead to a formation of primordial non-linear structures like stellar-mass-type black holes, and may include the SUSY GUTs inspired by heterotic string compactifications, unifying particle physics with quantum gravity.
\end{abstract}

dark energy, supersymmetry breaking,
primordial black holes


\newcommand{\overbar}[1]{\mkern1.5mu\overline{\mkern-1.5mu#1\mkern-1.5mu}\mkern 1.5mu}


\def\un#1{\relax\ifmmode\@@underline#1\else
        $\@@underline{\hbox{#1}}$\relax\fi}


\let\under=\unt                 
\let\ced=\ce                    
\let\du=\du                     
\let\um=\Hu                     
\let\sll=\lp                    
\let\Sll=\Lp                    
\let\slo=\os                    
\let\Slo=\Os                    
\let\tie=\ta                    
\let\br=\ub                     


\def\a{\alpha}
\def\b{\beta}
\def\c{\chi}
\def\d{\delta}
\def\e{\epsilon}
\def\f{\phi}
\def\g{\gamma}
\def\h{\eta}
\def\i{\iota}
\def\j{\psi}
\def\k{\kappa}
\def\l{\lambda}
\def\m{\mu}
\def\n{\nu}
\def\o{\omega}
\def\p{\pi}
\def\q{\theta}
\def\r{\rho}
\def\s{\sigma}
\def\t{\tau}
\def\u{\upsilon}
\def\x{\xi}
\def\z{\zeta}
\def\D{\Delta}
\def\F{\Phi}
\def\G{\Gamma}
\def\J{\Psi}
\def\L{\Lambda}
\def\O{\Omega}
\def\P{\Pi}
\def\Q{\Theta}
\def\S{\Sigma}
\def\U{\Upsilon}
\def\X{\Xi}


\def\ve{\varepsilon}
\def\vf{\varphi}
\def\vr{\varrho}
\def\vs{\varsigma}
\def\vq{\vartheta}


\def\ca{{\cal A}}
\def\cb{{\cal B}}
\def\cc{{\cal C}}
\def\cd{{\cal D}}
\def\ce{{\cal E}}
\def\cf{{\cal F}}
\def\cg{{\cal G}}
\def\ch{{\cal H}}
\def\ci{{\cal I}}
\def\cj{{\cal J}}
\def\ck{{\cal K}}
\def\cl{{\cal L}}
\def\cm{{\cal M}}
\def\cn{{\cal N}}
\def\co{{\cal O}}
\def\cp{{\cal P}}
\def\cq{{\cal Q}}
\def\car{{\cal R}}
\def\cs{{\cal S}}
\def\ct{{\cal T}}
\def\cu{{\cal U}}
\def\cv{{\cal V}}
\def\cw{{\cal W}}
\def\cx{{\cal X}}
\def\cy{{\cal Y}}
\def\cz{{\cal Z}}


\def\Sc#1{{\hbox{\sc #1}}}      
\def\Sf#1{{\hbox{\sf #1}}}      



\def\slpa{\slash{\pa}}                            
\def\slin{\SLLash{\in}}                                   
\def\bo{{\raise-.3ex\hbox{\large$\Box$}}}               
\def\cbo{\Sc [}                                         
\def\pa{\partial}                                       
\def\de{\nabla}                                         
\def\dell{\bigtriangledown}                             
\def\su{\sum}                                           
\def\pr{\prod}                                          
\def\iff{\leftrightarrow}                               
\def\conj{{\hbox{\large *}}}                            
\def\ltap{\raisebox{-.4ex}{\rlap{$\sim$}} \raisebox{.4ex}{$<$}}   
\def\gtap{\raisebox{-.4ex}{\rlap{$\sim$}} \raisebox{.4ex}{$>$}}   
\def\TH{{\raise.2ex\hbox{$\displaystyle \bigodot$}\mskip-4.7mu \llap H \;}}
\def\face{{\raise.2ex\hbox{$\displaystyle \bigodot$}\mskip-2.2mu \llap {$\ddot
        \smile$}}}                                      
\def\dg{\sp\dagger}                                     
\def\ddg{\sp\ddagger}                                   


\def\sp#1{{}^{#1}}                              
\def\sb#1{{}_{#1}}                              
\def\oldsl#1{\rlap/#1}                          
\def\slash#1{\rlap{\hbox{$\mskip 1 mu /$}}#1}      
\def\Slash#1{\rlap{\hbox{$\mskip 3 mu /$}}#1}      
\def\SLash#1{\rlap{\hbox{$\mskip 4.5 mu /$}}#1}    
\def\SLLash#1{\rlap{\hbox{$\mskip 6 mu /$}}#1}      
\def\PMMM#1{\rlap{\hbox{$\mskip 2 mu | $}}#1}   %
\def\PMM#1{\rlap{\hbox{$\mskip 4 mu ~ \mid $}}#1}       %
\def\Tilde#1{\widetilde{#1}}                    
\def\Hat#1{\widehat{#1}}                        
\def\Bar#1{\overline{#1}}                       
\def\sbar#1{\stackrel{*}{\Bar{#1}}}             
\def\bra#1{\left\langle #1\right|}              
\def\ket#1{\left| #1\right\rangle}              
\def\VEV#1{\left\langle #1\right\rangle}        
\def\abs#1{\left| #1\right|}                    
\def\leftrightarrowfill{$\mathsurround=0pt \mathord\leftarrow \mkern-6mu
        \cleaders\hbox{$\mkern-2mu \mathord- \mkern-2mu$}\hfill
        \mkern-6mu \mathord\rightarrow$}
\def\dvec#1{\vbox{\ialign{##\crcr
        \leftrightarrowfill\crcr\noalign{\kern-1pt\nointerlineskip}
        $\hfil\displaystyle{#1}\hfil$\crcr}}}           
\def\dt#1{{\buildrel {\hbox{\LARGE .}} \over {#1}}}     
\def\dtt#1{{\buildrel \bullet \over {#1}}}              
\def\der#1{{\pa \over \pa {#1}}}                
\def\fder#1{{\d \over \d {#1}}}                 
\def\Hat#1{\widehat{#1}}


\newskip\humongous \humongous=0pt plus 1000pt minus 1000pt
\def\caja{\mathsurround=0pt}
\def\eqalign#1{\,\vcenter{\openup2\jot \caja
        \ialign{\strut \hfil$\displaystyle{##}$&$
        \displaystyle{{}##}$\hfil\crcr#1\crcr}}\,}
\newif\ifdtup
\def\panorama{\global\dtuptrue \openup2\jot \caja
        \everycr{\noalign{\ifdtup \global\dtupfalse
        \vskip-\lineskiplimit \vskip\normallineskiplimit
        \else \penalty\interdisplaylinepenalty \fi}}}
\def\li#1{\panorama \tabskip=\humongous                         
        \halign to\displaywidth{\hfil$\displaystyle{##}$
        \tabskip=0pt&$\displaystyle{{}##}$\hfil
        \tabskip=\humongous&\llap{$##$}\tabskip=0pt
        \crcr#1\crcr}}
\def\eqalignnotwo#1{\panorama \tabskip=\humongous
        \halign to\displaywidth{\hfil$\displaystyle{##}$
        \tabskip=0pt&$\displaystyle{{}##}$
        \tabskip=0pt&$\displaystyle{{}##}$\hfil
        \tabskip=\humongous&\llap{$##$}\tabskip=0pt
        \crcr#1\crcr}}

 
 \newcommand{\be}{\begin{equation}}
\newcommand{\ee}{\end{equation}}
\newcommand{\nbe}{\begin{equation*}}
\newcommand{\nee}{\end{equation*}}

\newcommand{\fr}{\frac}
\newcommand{\lb}{\label}

\def\fracmm#1#2{{{#1}\over{#2}}}
  
\def\low#1{{\raise -3pt\hbox{${\hskip 0.75pt}\!_{#1}$}}}

\section{Introduction}\label{s:intro}

The Cosmic Microwave Background (CMB) data collected by the Planck collaboration \cite{Ade:2015xua, Ade:2015lrj,Array:2015xqh} favours the slow-roll single-field inflationary scenarios, with an approximately flat scalar potential.
The celebrated Starobinsky model \cite{Starobinsky:1980te} does provide such scenario, and relates its inflaton (called scalaron in this context) to the particular extension of Einstein-Hilbert gravity with the extra higher derivative term given by the scalar curvature squared, $ R^{2}$. However, a theoretical explanation of fundamental origin of  the Starobinsky model is still missing. The viable inflationary dynamics is driven by the $R^{2}$ term dominating over the (Einstein-Hilbert)  $R$ term. This is related to a missing UV completion of the  non-renormalizable $(R+ R^2)$ gravity.  The interesting and ambitious project for string phenomenology would be to provide a derivation of the Starobinsky model from the first principles.  A first step towards this is an embedding of the Starobinsky model  into four-dimensional $\mathcal{N}=1$ supergravity. In the supergravity framework,  the inflaton (scalaron) can mix with other scalars, and this mixing may ruin any initially successful inflationary mechanism. 
 
The inflationary model building based on supergravity in the literature usually assumes that inflaton belongs to a chiral (scalar) supermultiplet \cite{Ketov:2012yz,Ketov:2014qha,Ketov:2014hya}.  However, there is the alternative to this assumption: inflaton can also belong to a massive $\mathcal{N}=1$ vector multiplet.  The vector multiplet-based approach avoids stabilization problems related to the inflaton (scalar) superpartner, as the way-out of the standard $\eta$-problem. The scalar potential of a vector multiplet is given by the $D$-term instead of the $F$-term.  The minimal supergravity models, with inflaton belonging to a massive vector multiplet, were proposed in Refs.~\cite{Farakos:2013cqa,Ferrara:2013rsa}.  Then any desired values of the CMB observables (the scalar perturbations tilt $n_s$ and the 
tensor-to-scalar perturbations ratio  $r$) can be recast from the single-field (inflaton) scalar potential proportional to  the derivative squared of  arbitrary real function $J$. However, in these models, the vacuum energy is vanishing after inflation, thus restoring supersymmetry, and only a Minkowski vacuum is allowed.  The way-out of this problem was proposed in \cite{Aldabergenov:2016dcu,Aldabergenov:2017bjt} by adding a Polonyi (chiral) superfield with a linear superpotential \cite{Polonyi:1977pj}, leading to a spontaneous SUSY breaking and allowing a de-Sitter vacuum after inflation.

A successful model of inflation in supergravity should also be consistent with the Cold Dark Matter (CDM) constraints and the Big Bang Nucleosynthesis (BBN).  For example, many supergravity scenarios are plagued by the so-called gravitino problem. Gravitinos can decay, injecting hadrons and photons during the BBN epoch, which may jeopardize the good Standard Model prediction of nuclei ratios \cite{Khlopov:1984pf,Khlopov:1993ye,Kawasaki:2004qu,Khlopov:2015oda}. In very much the same way, the Polonyi (overproduction) problem and its relation to the BBN results were extensively discussed in the literature \cite{Banks:1993en,deCarlos:1993wie,Coughlan:1983ci,Moroi:1994rs,Kawasaki:1995cy,Moroi:1999zb}. In addressing these issues, the mass spectrum and the soft SUSY parameters are important. The leading (WIMP-like) dark matter production mechanisms and decay channels are selected from the mass pattern, and have either thermal or non-thermal origin. 

In this paper, we review a class of the minimalistic Polonyi-Starobinsky (PS) $\mathcal{N}=1$ supergravity models for inflation, with the inflaton belonging to a (massive) vector multiplet. These models can avoid the overproduction and BBN problems,  while accounting for the right amount of CDM composed of gravitinos.  In our analysis, we assume that the Polonyi field, inducing a spontaneous SUSY breaking at a high energy scale, and the gravitino, as the Dark Matter (DM) particle, are both super-heavy. The main mechanism producing DM is given by the Schwinger-type production sourced by inflationary expansion.  After inflation, Polonyi particles  rapidly decay into gravitinos.  We find 
 that gravitinos produced directly  from Schwinger's production and  from Polonyi particles decays, can account for the correct abundance of Cold Dark Matter. 
 
Another aspect  is an inclusion of the (mini) Primordial Black Holes (PBHs) that may have been copiously produced in the early Universe, and later may have evaporated into gravitinos and other Standard Model particles \cite{Khlopov:1985jw,Khlopov:2004tn,Khlopov:2008qy,miniPBH,miniPBH2}.  A large amount of mini PBHs cannot be produced  in our model when the other scalar and pseudo-scalar partners of inflaton  are not participating in the inflationary dynamics. The Starobinsky inflaton entails a scalar potential shape that cannot lead to a large number of PBHs, because it does not allow for amplifying instabilities and has no exit out of inflation with a first order phase transition. It is still possible that dynamics of other scalar fields changes this picture. In this case, the extra moduli can exit from inflation via ending in false minima. The tunneling process from a false minimum to the true one sources the production of bubbles related to the first order phase transition. 

As regards the (solar mass type) PBHs, their production in the early Universe is possible in our supergravity approach after a certain
deformation of the Starobinsky scalar potential. We envisage a unification of the inflaton in a vector multiplet and the Supersymmetric Grand Unified Theories (SUSY GUTs), whose gauge group has at least one abelian factor, such as the flipped SU(5)$\times$U(1) model arising from the compactified heterotic superstrings or the intersecting D-branes.

\section{Starobinsky model of $(R+R^2)$ gravity}

Starobinsky model of inflation is defined by the action \cite{Starobinsky:1980te}
 \be \label{star}
S_{\rm Star.} = \fracmm{M^2_{\rm Pl}}{2}\int \mathrm{d}^4x\sqrt{-g} \left( R +\fracmm{1}{6m^2}R^2\right)~,
\ee
where we have introduced the reduced Planck mass $M_{\rm Pl}=1/\sqrt{8\p G_{\rm N}}\approx
2.4\times 10^{18}$ GeV, and the scalaron (inflaton) mass $m$ as the only parameter. We use the
spacetime signature $(-,+,+,+,)$. The $(R+R^2)$ gravity model (\ref{star}) can be considered as the simplest extension of the standard Einstein-Hilbert action in the context of (modified) $F(R)$ gravity theories with an action
 \be \label{fg}
S_F = \fracmm{M^2_{\rm Pl}}{2}\int \mathrm{d}^4x\sqrt{-g} \, F(R)~,
\ee
in terms of the function $F(R)$ of the scalar curvature $R$.

The $F(R)$ gravity action (\ref{fg}) is classically equivalent to 
\begin{equation} \lb{eq}
 S[g_{\m\n},\chi] = \fracmm{M^2_{\rm Pl}}{2} \int d^{4}x \sqrt{- g}~ \left[ F'(\chi) (R - \chi) + F(\chi) \right]
\end{equation}
with the real scalar field $\chi$, provided that $F''\neq 0$ that we always assume.  Here the
primes denote the derivatives with respect to the argument. The equivalence is easy to verify because the $\chi$-field equation implies $\chi=R$. In turn, the factor $F'$ in front of the $R$ in (\ref{eq}) can be (generically) eliminated by a Weyl transformation of metric $g_{\m\n}$, that transforms the action (\ref{eq}) into the action of the scalar field   $\chi$ minimally coupled to Einstein gravity and having the scalar potential 
\be \lb{spot}
 V =      \left(\fracmm{M^2_{\rm Pl}}{2}\right)       \fracmm{\chi F'(\chi) - F(\chi)}{F'(\chi)^{2}}~~.
\ee
Differentiating this scalar potential yields
\be \lb{diffp}
\frac{dV}{d\chi} =  \left(\fracmm{M^2_{\rm Pl}}{2}\right)    
\fracmm{F''(\chi)\left[2F(\chi) - \chi F'(\chi)\right]}{ (F'(\chi))^{3}}~~.
\ee

The kinetic term of $\chi$ becomes canonically normalized after the field redefinition $\chi(\varphi)$ as
\begin{equation} \lb{fred}
  F'(\chi) =  \exp \left(  \sqrt{\frac{2}{3}} \varphi/M_{\rm Pl} \right)~,\quad
  \varphi =  \fracmm{\sqrt{3}M_{\rm Pl}}{\sqrt{2}}\ln F'(\c) ~~,
\end{equation}
in terms of the canonical inflaton field $\varphi$, with the total acton
\be \lb{quint}
S_{\rm quintessence}[g_{\m\n},\varphi]  = \fracmm{M^2_{\rm Pl}}{2}\int \mathrm{d}^4x\sqrt{-g} R 
 - \int \mathrm{d}^4x \sqrt{-g} \left[ \frac{1}{2}g^{\m\n}\pa_{\m}\varphi\pa_{\n}\varphi
 + V(\varphi)\right]~.
\ee

The classical and quantum stability conditions of $F(R)$ gravity theory are given by 
\cite{Ketov:2012yz}
\begin{equation} \lb{stab}
 F'(R) > 0 \quad {\rm and} \quad F''(R) > 0~,
\end{equation}
and they are obviously satisfied for Starobinsky model (\ref{star}) for $R>0$.

Differentiating the scalar potential $V$ in Eq.~(\ref{spot}) with respect to $\varphi$ yields 
\begin{equation} \lb{diff1}
    \frac{d V}{d \varphi}   = \frac{d V}{d \chi} \frac{d \chi}{d \varphi} 
      = \frac{M^2_{\rm Pl}}{2} \left[ \frac{\chi F'' + F' - F'}{F'^{2}} - 2 \frac{\chi F' - F}{F'^{3}} F'' \right] 
\frac{d \chi}{d \varphi}~~,
\end{equation}
where we have
\begin{equation} \lb{diff2}
  \frac{d \chi}{d \varphi}
    = \frac{d \chi}{d F'} \frac{ d F'}{d \varphi}
    = \frac{d F'}{d \varphi} \left/ \frac{d F'}{d\chi} \right.
    = \fracmm{\sqrt{2}}{\sqrt{3}M_{\rm Pl}}  \frac{F'}{F''}~~.
\end{equation}
This implies
\begin{equation} \lb{derv}
  \frac{d V}{d \varphi} = M_{\rm Pl}\frac{2F-\chi F'}{\sqrt{6}  F'^{2}}~~.
\end{equation}
Combining Eqs.~(\ref{spot}) and (\ref{derv})  yields $R$ and $F$ in terms of the scalar potential $V$,
\begin{align} \lb{inv}
  & R = \left[  \fracmm{\sqrt{6}}{M_{\rm Pl}}
    \frac{d V}{d \varphi} + \fracmm{4V}{M^2_{\rm Pl}} \right] \exp \left(  \sqrt{\frac{2}{3}} 
  \varphi/M_{\rm Pl} \right),  \\
  & F= \left[  \fracmm{\sqrt{6}}{M_{\rm Pl}}
 \frac{d V}{d \varphi} + \fracmm{2V}{M^2_{\rm Pl}} \right] \exp \left(  2 \sqrt{\frac{2}{3}} \varphi/M_{\rm Pl}\right).
\end{align}
These equations define the function $F(R)$ in the parametric form, in terms of a scalar potential $V(\varphi)$, i.e. the {\it inverse} transformation to (\ref{spot}). This is known \cite{mf}  as the classical equivalence (duality) between the $F(R)$ gravity theories (\ref{fg}) and the scalar-tensor (quintessence) theories of gravity (\ref{quint}).

In the case of Starobinsky model (\ref{star}), one gets the famous potential
\begin{equation} \label{starp}
V(\varphi) = \fracmm{3}{4} M^2_{\rm Pl}m^2\left[ 1- \exp\left(-\sqrt{\frac{2}{3}}\varphi/M_{\rm Pl}\right)\right]^2~.
\end{equation}
This scalar potential is bounded from below (non-negative and stable), and it has the absolute  minimum at 
$\varphi=0$  corresponding to  a Minkowski vacuum. The scalar potential (\ref{starp}) also has a {\it plateau} of positive height (related to inflationary energy density), that gives rise to slow roll of inflaton in the inflationary era.
 The Starobinsky model (\ref{star}) is the particular case of the so-called $\alpha$-attractor inflationary models \cite{Galante:2014ifa}, and is also a member of the close family of viable inflationary models of $F(R)$ gravity, originating from higher dimensions \cite{Nakada:2017uka}.

A duration of inflation is measured in the slow roll approximation by the e-foldings number
\be \lb{efold}
N_e\approx  \fracmm{1}{M^2_{\rm Pl}} \int_{\varphi_{\rm end}}^{\varphi_{*}} \fracmm{V}{V'}d\varphi~~,
\ee
where $\varphi_{*}$ is the inflaton value at the reference scale (horizon crossing), and $\varphi_{\rm end}$ is the
inflaton value at the end of inflation when one of the slow roll parameters
\be \lb{slowp}
\ve_V(\varphi) = \fracmm{M^2_{\rm Pl}}{2}\left( \fracmm{V'}{V}\right)^2 \quad {\rm and} \quad 
\eta_V(\varphi) = M^2_{\rm Pl} \left( \fracmm{V''}{V}\right)~~,
\ee
is no longer small (close to 1).

The amplitude of scalar perturbations at horizon crossing is given by \cite{Ellis:2015pla}
\be \lb{amp}
A = \fracmm{V_*^3}{12\p^2 M^6_{\rm Pl}({V_*}')^2}=\fracmm{3m^2}{8\p^2M^2_{\rm Pl}}\sinh^4\left(
\fracmm{\varphi_*}{\sqrt{6}M_{\rm Pl}}\right)~~.
\ee

The Starobinsky model  (\ref{star}) is the excellent model of cosmological inflation, in very good agreement with the Planck data \cite{Ade:2015xua,Ade:2015lrj,Array:2015xqh}. The Planck satellite mission measurements of the Cosmic Microwave Background (CMB) radiation \cite{Ade:2015xua,Ade:2015lrj,Array:2015xqh}
give the scalar perturbations tilt as $n_s\approx 1+2\eta_V -6\ve_V\approx 0.968\pm 0.006$ and restrict the 
tensor-to-scalar ratio as $r\approx 16\ve_V < 0.08$. The Starobinsky inflation yields $r\approx 12/N_e^2\approx 0.004$ and $n_s\approx 1- 2/N_e$, where $N_e$ is the e-foldings number between 50 and 60, with the best fit at $N_e\approx 55$ \cite{Mukhanov:1981xt,Kaneda:2010ut}.

The Starobinsky model (\ref{star}) is geometrical (based on gravity only), while its (mass) parameter $m$ is fixed by the 
observed CMB amplitude (COBE, WMAP) as 
\be \lb{starm}
m\approx 3 \cdot10^{13}~{\rm GeV} \quad {\rm or}\quad \fracmm{m}{M_{\rm Pl}}\approx 1.3\cdot 10^{-5}~.
\ee
A numerical analysis of (\ref{efold}) with the potential (\ref{starp}) yields \cite{Ellis:2015pla}
\be\lb{sandf}
\sqrt{\fracmm{2}{3}} \varphi_*/M_{\rm Pl} \approx \ln\left( \fracmm{4}{3}N_e\right) \approx 5.5~,
\quad \sqrt{\fracmm{2}{3}} \varphi_{\rm end}/M_{\rm Pl} \approx \ln\left[ \fracmm{2}{11}(4+3\sqrt{3})\right]\approx 0.5~~,
\ee
where $N_e\approx 55$ has been used.

\section{Starobinsky inflation in supergravity}

Let us introduce a set of two chiral superfields $(\Phi, H)$ and a real vector superfield $V$ coupled to the supergravity sector,
with the following Lagrangian:~\footnote{We use the standard notation \cite{Wess:1992cp}  for supergravity in superspace.}
\begin{equation}
\label{hsslag}
\mathcal{L}=\int d^2\theta 2\mathcal{E}\left\lbrace \frac{3}{8}(\overbar{\mathcal{D}}\overbar{\mathcal{D}}-8\mathcal{R})e^{-\frac{1}{3}(K+2J)}+\frac{1}{4}W^\alpha W_\alpha +\mathcal{W}(\Phi) 
\right\rbrace +{\rm h.c.}~,
\end{equation}
where $\mathcal{R}$ is the chiral scalar curvature superfield,  $\mathcal{E}$ is the chiral density superfield,
 $(\mathcal{D}_{\alpha}, \overbar{\mathcal{D}}^{\dt{\alpha}})$ are the superspace covariant spinor derivatives,  
 $K= K(\Phi,\overbar{\Phi})$ is the K\"ahler potential, 
$\mathcal{W}(\Phi)$ is the superpotential, 
$W_\alpha\equiv-\frac{1}{4}(\overbar{\mathcal{D}}\overbar{\mathcal{D}}-8\mathcal{R})\mathcal{D}_\alpha V$
is the abelian (chiral) superfield strength, and $J=J(He^{2gV}\overbar{H})$
is a real function with the coupling constant $g$.
   
The Lagrangian \eqref{hsslag} is invariant under the supersymmetric $U(1)$ gauge transformations
\begin{gather}
H\rightarrow H'=e^{-igZ}H~,\;\;\;\overbar{H}\rightarrow \overbar{H}'=e^{ig\overbar{Z}}\overbar{H}~,\label{supergh}\\
V\rightarrow V'=V+\frac{i}{2}(Z-\overbar{Z})~,\label{supergv}
\end{gather}
the gauge parameter of which, $Z$, is itself a chiral superfield. The chiral superfield $H$ can be gauged away via the gauge fixing of these transformations by imposing the gauge condition $H=1$. Then the Lagrangian \eqref{hsslag} gets simplified to
\begin{equation} \label{sslag}
\mathcal{L}=\int d^2\theta 2\mathcal{E}\left\lbrace \frac{3}{8}(\overbar{\mathcal{D}}\overbar{\mathcal{D}}-8\mathcal{R})e^{-\frac{1}{3}(K+2J)}
+\frac{1}{4}W^\alpha W_\alpha +\mathcal{W} \right\rbrace +{\rm h.c.}
\end{equation}

After eliminating the auxiliary fields and moving from the initial (Jordan) frame to the Einstein frame, the {\it bosonic} part of the Lagrangian  (\ref{sslag}) reads \cite{Aldabergenov:2016dcu}~\footnote{The primes and capital latin subscripts denote the derivatives with respect to the corresponding fields.} 
 \begin{equation} \label{complag}
e^{-1}\mathcal{L}=-\frac{1}{2}R-K_{A\bar{A}}\partial_m A\partial^m\bar{A}-\frac{1}{4}F_{mn}F^{mn}-\frac{1}{2}J''\partial_mC\partial^mC-\frac{1}{2}J''B_mB^m-\mathcal{V}~,
\end{equation}
with the scalar potential
\begin{equation} \label{pot}
\mathcal{V}=\fracmm{g^2}{2}{J'}^2+e^{K+2J}{}\left\{
	K^{-1}_{A\bar{A}}(\mathcal{W}_A+K_A\mathcal{W})(\overbar{\mathcal{W}}_{\bar{A}}+K_{\bar{A}}\overbar{\mathcal{W}})-\bigg(3-2\fracmm{{J'}^2}{J''}\bigg)\mathcal{W}\overbar{\mathcal{W}}
	\right\}
\end{equation}
in terms of the physical fields ($A$, $C$, $B_m$), the auxiliary fields ($F$, $X$, $D$) and the vector field strength 
$F_{mn}=\mathcal{D}_mB_n-\mathcal{D}_nB_m$. 

As is clear from Eq.~\eqref{complag}, the absence of ghosts requires $J''(C)>0$, where the primes denote the differentiations with respect to the given argument. We restrict ourselves to the 
K\"ahler potential and the superpotential of the {\it Polonyi model} \cite{Polonyi:1977pj}:
\begin{equation} \label{polonyi}
K= \Phi\overbar{\Phi}~,\qquad \mathcal{W}=\mu(\Phi +\beta)~,
\end{equation}
with the parameters $\mu$ and $\beta$.  Our model includes the single-field $(C)$ inflationary model, whose $D$-type scalar potential is  given by 
\begin{equation} \label{scalpot}
V(C)=\frac{g^2}{2}({J'})^2
\end{equation}
in terms of {\it arbitrary} function $J(C)$, with the real inflaton field $C$ belonging to a massive vector supermultiplet.
The Minkowski vacuum conditions (after inflation) can be easily satisfied when $J'=0$, which implies \cite{Polonyi:1977pj}
 \begin{equation} \label{vevs}
 \VEV{A}=\sqrt{3}-1 \qquad {\rm and} \qquad \beta=2-\sqrt{3}~~.
 \end{equation}
This solution describes a {\it stable} Minkowski vacuum with spontaneous SUSY breaking at {\it arbitrary} scale 
$\langle F\rangle=\mu$. The related gravitino mass is given by 
\begin{equation} \label{gmass}
m_{3/2}=\mu e^{2-\sqrt{3}+\VEV{J}}~~.
\end{equation}
There is also a complex (Polonyi) scalar of mass 
\begin{equation} \label{massrel}
M_A  = 2 \mu e^{2-\sqrt{3}} \geq 2m_{3/2}
\end{equation}
 and a massless fermion in the physical spectrum. The inequality in Eq.~(\ref{massrel}) is saturated in the original
 Polonyi model \cite{Polonyi:1977pj} but it is not the case in our model when $\VEV{J}<0$.
 
As regards the early Universe phenomenology, our model has the following theoretically appealing features:
\begin{itemize}
\item there is no need to "stabilize" the single-field inflationary trajectory against scalar superpartners
of inflaton, because our inflaton is the only real scalar in a massive vector multiplet,
\item any values of CMB observables $n_s$ and $r$ are possible by choosing the $J$-function,
\item a spontaneous SUSY breaking after inflation occurs at arbitrary scale $\mu$,
\item there are only {\it a few} parameters relevant for inflation and SUSY breaking: the coupling constant $g$ defining the inflaton mass, $g\sim m_{\rm inf.}$, the coupling constant $\mu$ defining the scale of SUSY breaking, $\mu\sim m_{3/2}$, and the parameter $\beta$ in the constant term of the superpotential. Actually,  the inflaton mass is constrained by CMB observations as $m_{\it inf.}\sim {\cal O}(10^{-6})$, while $\beta$ is fixed by the vacuum  solution,  so that we have only {\it one} free parameter $\mu$
defining the scale of SUSY breaking in our model (before studying reheating and phenomenology).
\end{itemize}

The D-type scalar potential associated with the Starobinsky inflationary model of $(R+R^2)$ gravity arises when 
\cite{Ferrara:2013rsa}
\begin{equation} \label{starj}
 J(C)=  \frac{3}{2} \left( C- \ln C\right)  
 \end{equation}
that implies
\begin{equation} \label{starjder}
 J'(C) =   \frac{3}{2}\left(1- C^{-1}\right)  \qquad {\rm and} \qquad  J'' (C)=   \frac{3}{2}\left(C^{-2}\right)>0~.
 \end{equation}
According to  \eqref{complag}, a canonical inflaton field $\phi$ (with the canonical kinetic term) is related to the field $C$ by the field redefinition
\begin{equation} \label{canonf}
C =  \exp\left( \sqrt{2/3} \phi\right)~.
\end{equation}
Therefore, we arrive at the (Starobinsky) scalar potential 
\begin{equation} \label{starpot}
V_{\rm Star.}(\phi) = \frac{9g^2}{8}  \left( 1- e^{-\sqrt{2/3}\phi}   \right)^2\quad {\rm with} \quad
m^2_{\it inf.}=9g^2/2~~. 
\end{equation}

The full action \eqref{hsslag} of this PS supergravity in curved superspace can be transformed into a supergravity extension of the $(R+R^2)$ gravity action  by using the (inverse) duality procedure described in Ref.~\cite{Ferrara:2013rsa}.  However, the dual supergravity model is described by a  complicated {\it higher-derivative}  field theory that is inconvenient for studying particle production.

Another nice feature of our model is that it can be rewritten as a supersymmetric (abelian and non-minimal) gauge theory coupled to supergravity in the presence of a {\it Higgs} superfield $H$, resulting in the super-Higgs effect with simultaneous spontaneous breaking of the gauge symmetry and SUSY. Indeed, the $U(1)$ gauge symmetry of the original Lagrangian (\ref{hsslag}) allows us to choose a different ({\it Wess-Zumino}) supersymmetric gauge by "gauging away" the chiral and anti-chiral parts of the general superfield $V$ via the appropriate choice of the superfield parameters $Z$ and $\overbar{Z}$. Then the bosonic part of the Lagrangian in terms of the superfield components in the Einstein frame, after elimination of the auxiliary fields and Weyl rescaling, reads \cite{Aldabergenov:2017bjt}
\begin{multline} \label{dhlag2}
e^{-1}\mathcal{L}=-\frac{1}{2}R-K_{AA^*}\partial^m A\partial_m\bar{A}-\frac{1}{4}F_{mn}F^{mn}-2J_{h\bar{h}}\partial_mh\partial^m\bar{h}-\frac{1}{2}J_{V^2}B_mB^m\\+iB_m(J_{Vh}\partial^mh-J_{V\bar{h}}\partial^m\bar{h})-\mathcal{V}~,
\end{multline}
where $h$, $\bar{h}$ are the Higgs field and its conjugate.

The standard $U(1)$ Higgs mechanism arises with the canonical function $J=\frac{1}{2}he^{2V}\bar{h}$, where we
have chosen $g=1$ for simplicity.  As regards the Higgs sector, it leads to 
\begin{equation}
e^{-1}\mathcal{L}_{Higgs}=-\partial_mh\partial^m\bar{h}+iB_m(\bar{h}\partial^mh-h\partial^m\bar{h})-h\bar{h}B_mB^m-\mathcal{V}~.
\end{equation}
After changing the variables $h$ and $\bar{h}$ as
\begin{equation}
h=\frac{1}{\sqrt{2}}(\rho+\nu)e^{i\zeta},\;\;\;\bar{h}=\frac{1}{\sqrt{2}}(\rho+\nu)e^{-i\zeta}~,\label{paramh}
\end{equation}
where $\rho$ is the (real) Higgs boson, $\nu\equiv \langle h\rangle=\langle \bar{h}\rangle$ is the Higgs VEV, and $\zeta$ is the Goldstone boson,  the unitary gauge fixing of $h\rightarrow h'=e^{-i\zeta}h$ and $B_m\rightarrow B'_m=B_m+\partial_m\zeta$, leads to the standard result 
\begin{equation}
e^{-1}\mathcal{L}_{Higgs}=-\frac{1}{2}\partial_m\rho\partial^m\rho-\frac{1}{2}(\rho+\nu)^2B_mB^m-\mathcal{V}~.
\end{equation}

The Minkowski vacuum after inflation can be easily lifted to a {\it de Sitter} vacuum (Dark Energy) in our model by the simple modification of the Polonyi sector and its parameters as \cite{Aldabergenov:2017bjt}
\begin{equation}
\VEV{A}=(\sqrt{3}-1)+\fracmm{3-2\sqrt{3}}{3(\sqrt{3}-1)}\delta+\mathcal{O}(\delta^2)~,\quad \beta=(2-\sqrt{3})+\fracmm{\sqrt{3}-3}{6(\sqrt{3}-1)}\delta+\mathcal{O}(\delta^2)~,
\end{equation}
where $\delta$ is a very small deformation parameter, $0<\delta\ll 1$. It leads to a positive cosmological constant
\begin{equation}
V_0=\mu^2e^{\alpha^2}\delta=m^2_{3/2}\delta\label{dsvac}
\end{equation}
and the superpotential  VEV
\begin{equation}
\langle \mathcal{W}\rangle=\mu(\VEV{A} +\beta)=\mu(a+b-\frac{1}{2}\delta)~,
\end{equation}
where $a\equiv(\sqrt{3}-1)$ and $b\equiv(2-\sqrt{3})$ provide the SUSY breaking vacuum solution to the Polonyi parameters in the absence of a cosmological constant.

The full scalar potential (\ref{pot}) is a sum of the D- and F-type terms, while there is a mix of the inflaton - and Polonyi-dependent terms in the F-type contribution. This mixing leads to instability of the (Starobinsky) inflationary trajectory that is supposed to be driven by the D-term only. This issue was resolved in Ref.~\cite{Aldabergenov:2017hvp} where a modification of the original PS supergravity action (\ref{hsslag}) was proposed via adding the generalized Fayet-Iliopoulos term and modifying the $J$-function (\ref{starj}).

\section{Super heavy gravitino dark matter}

The complete set of equations of motion in our supergravity model (Sec.~3) is very complicated. In this section, we consider only the leading order with respect to the inverse Planck mass.  In addition, we neglect the coupling of Polonyi and gravitino particles to the inflaton, and  introduce the effective action of the Polonyi field in the Friedmann-Lemaitre-Robertson-Walker (FLRW) background (in comoving coordinates) as
\begin{equation}
\label{Sdt}
I[A]=\int dt \int d^{3}x \fracmm{a^{3}}{2}\left(\dot{A}^{2}-\fracmm{1}{a^{2}}(\nabla A)^{2}-M_{A}^{2}A^{2}-\zeta R A^{2}\right)~,
\end{equation}
where the non-minimal coupling constant of the Polonyi field  to gravity is equal to $\zeta=1$, $A$ is the Polonyi field, $M_{A}$ stands for its mass, $R$ is the Ricci scalar, and $a$ is the FLRW scale factor. 

The mode decomposition of the  Polonyi field reads 
\begin{equation}
\label{XX}
A({\bf x})=\int d^{3}k (2\pi)^{-3/2}a^{-1}(\eta)\left[ b_{k}h_{k}(\eta)e^{i{\bf k}\cdot {\bf x}}+b_{k}^{\dagger}h^{*}_{k}(\eta)e^{-i{\bf k}\cdot {\bf x}}\right]\, ,
\end{equation}
where the conformal time coordinate $\eta$ is introduced, 
 $b,b^{\dagger}$  are the (standard)  creation/annihilation operators, and the 
coefficient functions $h,h^{+}$ are normalized as follows: 
\begin{equation}\label{hh}
h_{k}h'^{*}_{k}-h'_{k}h^{*}_{k}=i~~.
\end{equation}

Because of Eqs.~(\ref{Sdt}) and (\ref{XX}),  the equation of motion of the modes is 
\begin{equation}
\label{modes}
h''_{k}(\eta)+\omega_{k}^{2}(\eta)h_{k}(\eta)=0~, \quad { \rm where}  \quad \omega_{k}^{2}=5\fracmm{a''}{a}+k^{2}+M_{A}^{2}a^{2}~~,
\end{equation}
and $h''=d^{2} h/d\eta^{2}$. Equation (\ref{modes})  can be conveniently rescaled by  using some reference scales  
$a(\eta_{*})\equiv a_*$ and $H(\eta_{*})=H_{*}$  as follows:
\begin{equation}
\label{res}
h''_{\tilde{k}}(\tilde{\eta})+(\tilde{k}^{2}+b^{2}\tilde{a}^{2})h_{\tilde{k}}(\tilde{\eta})=0\, ,
\end{equation}
in terms of the rescaled quantities
$$\tilde{\eta}=\eta a_{*}H_{*}~,\quad \tilde{a}=a/a_{*}~,\quad \tilde{k}=k/(H_{*}a_{*})~~.$$

The leading order of the gravitino action coincides with the massive Rarita-Schwinger action,
\begin{equation}
\label{Rarita}
I[\psi]=\int d^{4}x \,e\, \bar{\psi}_{\sigma}\mathcal{R}^{\sigma}\{\psi\}\, ,
\end{equation}
where the gravitino kinetic operator has been introduced as
\begin{equation}
\label{RC}
\mathcal{R}^{\sigma}\{\psi\}=m_{3/2}\gamma^{\sigma\nu}\psi_{\nu}+i \gamma^{\sigma\nu\rho}\mathcal{D}_{\nu} \psi_{\rho} ~,
\end{equation}
and the supercovariant derivative is
\begin{equation}
\label{cov}
\mathcal{D}_{\mu}\psi_{\nu}=-\Gamma_{\mu\nu}^{\rho}\psi_{\rho}+\partial_{\mu}\psi_{\nu}+\frac{1}{4}\omega_{\mu ab}\gamma^{ab}\psi_{\nu}~~,
\end{equation}
in the $\gamma$-notation $\gamma^{\mu_{1}...\mu_{n}}=\gamma^{[\mu_{1}}....\gamma^{\mu_{n}]}$.

Since  the supergravity torsion is of the second order with respect to the inverse Planck mass, we ignore it in the leading order approximation. The $\Gamma_{\mu\nu}^{\rho}$ can be represented by the standard symmetric Christoffel symbols that are actually cancelled from the Rarita-Schwinger action (\ref{Rarita}). The Rarita-Schwinger action leads to the gravitino equation of motion, 
\begin{equation}
\label{eqRSS}
(i  \mathcal{\slashed{D}}-m_{3/2})\psi_{\mu}-\left(i\mathcal{D}_{\mu}+\frac{m_{3/2}}{2}\gamma_{\mu}\right)\gamma \cdot \psi=0~~.
\end{equation}
In the flat FLRW background,
Eq.~(\ref{eqRSS}) reduces to   
\begin{equation}
\label{psiEOMS}
i\gamma^{mn}\partial_{m}\psi_{n}=-\left(m_{3/2}+i\fracmm{a'}{a}\gamma^{0}\right)\gamma^{m}\partial_{m}\psi~~,
\end{equation}
where 
\begin{equation}
\label{cond}
\omega_{\mu ab}=2\dot{a}a^{-1}e_{\mu[a}e_{b]}^{0}~,\quad e_{\mu}^{a}=a(\eta) \delta_{\mu}^{a}~,\quad m_{3/2}=m_{3/2}(\eta)~.
\end{equation}
A solution to Eq.~(\ref{psiEOMS}) is 
\begin{equation}
\label{grav}
\psi_{\mu}(x)=\int d^{3}{\bf p}(2\pi)^{-3}(2p_{0})^{-1}\sum_{\lambda}\{ e^{i{\bf k}\cdot {\bf x}}b_{\mu}(\eta,\lambda)a_{k\lambda}(\eta)+e^{-i{\bf k}\cdot {\bf x}}b_{\mu}^{C}(\eta,\lambda)a_{k\lambda}^{\dagger}(\eta) \}~~.
\ee

We find that the equations of motion for the $3/2$-helicity gravitino modes have the same form as that of 
Eq.~(\ref{modes}), namely,
\begin{equation}
\label{Modes1}
b''_{\mu}(\eta,\lambda)+\hat{C}(k,a)b'_{\mu}(\eta,\lambda)+\omega^{2}(k,a)b_{\mu}(\eta,\lambda)=0~,
\end{equation}
where we have introduced the notation 
\begin{equation}
\label{Ca}
\hat{C}(k,a)b'_{\mu}(\eta,\lambda)=-2i\gamma^{\nu i}k_{i}\gamma_{\nu\eta}\partial^{\eta}b_{\mu}-2\gamma_{\nu}(m_{3/2}+i\frac{a'}{a}\gamma^{0})i\gamma^{\nu\eta}\partial_{\eta}b_{\mu}~~,
\end{equation}
\begin{equation}
\label{omega2}
\omega^{2}(k,a)/2= k^{2}+m_{3/2}^{2}+2i\frac{a'}{a}\gamma^{0}m_{3/2}
-\left(\frac{a'}{a}\right)^{2}~~.
\end{equation}
Following a procedure similar to the standard one in the case of  Dirac  and Klein-Gordon equations,  
we can reformulate the mode equations of motion in our case as
\begin{equation}
\label{PP}
P_{\nu}P^{\nu}b_{\mu}(\eta,\lambda)=0~~,
\end{equation}
where we have introduced the projector operator 
\begin{equation}
\label{PPP}
P^{\nu}=i\gamma^{\nu\eta}\partial_{\eta}-\gamma^{\nu i}k_{i}-\left(m_{3/2}+i\frac{a'}{a}\gamma^{0} \right)\gamma^{\nu}=0\, . 
\end{equation}

The dynamics of the gravitino and Polonyi fields during inflation necessary lead to their quantum production. The number density of produced particles can be calculated by using a Bogoliubov transformation, 
\begin{equation}
\label{BOGO}
h_{k}^{\eta_{1}}(\eta)=\alpha_{k}h_{k}^{\eta_{0}}(\eta)+\beta_{k}h_{k}^{*\eta_{0}}(\eta)\, .
\end{equation}
This transformation is performed from the vacuum solution selected by the boundary conditions at $\eta=\eta_{in}$, corresponding to the initial time of inflation, to the final time $\eta=\eta_{f}$,  when the particles creations process from inflation stops. In the inflationary epoch, the dynamical regime is $a'/a^{2}\ll M_{Pl}$ and $M_{Pl}\,ba/k\ll 1$. This implies that we can consider the extremes as $\eta_{in}=-\infty$ and $\eta_{f}=+\infty$, performing a WKB semiclassical approximation. By assuming these boundary conditions, the energy density of the Polonyi particles produced during inflation reads 
\begin{equation}
\label{S}
\rho_{A}(\eta_{})=M_{A}n_{A}(\eta_{})=M_{A}H_{\rm inf}^{3}\left(\fracmm{1}{\tilde{a}(\eta_{})} \right)^{3}\mathcal{P}_{A}~~ ,
\end{equation}
where 
\begin{equation}
\label{power}
\mathcal{P}_{A}=\fracmm{1}{2\pi^{2}}\int_{0}^{\infty}d\tilde{k}\tilde{k}^{2}|\beta_{\tilde{k}}|^{2}~~. 
\end{equation}

The inflaton mass sets the characteristic energy scale for the Hubble constant, calculated at fixed cosmological time $t\equiv t_{f}$:
$$H^{2}(t_{f})\simeq m_{\phi}^{2},\,\,\,\rho(t_{f})\simeq m_{\phi}^{2}M_{Pl}^{2}\, .$$

We propose the following formula for Polonyi particles (energy-density and Polonyi mass) produced during inflation \cite{Addazi:2017ulg}:
\begin{equation}
\label{rewe}
(\Omega_{A}h^{2}/\Omega_{R}h^{2})\simeq \fracmm{8\pi}{3}\left( \fracmm{M_{A} }{M_{Pl}}\right)\left(\fracmm{T_{\rm reh}}{T_{0}} \right)\fracmm{n_{A}(t_{f})}{M_{Pl}H^{2}(t_{f})}~~,
\end{equation}
where $M_{A}$ is the Polonyi mass, $\Omega_{R}h^2\simeq 4.31 \times 10^{-5}$ is the radiation energy density at today's temperature $T_0$, $\Omega_{A}h^2$ is the energy density of the produced Polonyi fields, all in the units of the critical energy density.  There is about $8th$-orders-of-magnitude suppression of the energy density.  The normalized power spectrum  $\mathcal{P}_{A}$ cannot provide such suppression  with our values for $M_A$ and $H_{inf}$. However, it comes from the dilution factor $(\tilde{a})^{-3}=(a_{f}/a_{i})^{-3}$ in Eq.~(\ref{S}). 
 
To get the gravitino and Polonyi masses, we have to add a few cosmological assumptions about the relevant parameters of the reheating process and, in particular, about the reheating temperature $T_{\rm reh}$. The cosmological parameters can be fixed by specifying the 
e-foldings number $N_{e}$ in the range between $50$ and $60$. For a more precise estimate of the CDM abundance, 
we choose $N_{e}=55$, as in Sec.~2. This implies $n_{s}=0.964$, $r=0.004$, $m_{inf}=3.2\cdot 10^{13}\, {\rm GeV}$ and
$ H_{inf}=\p M_P\sqrt{P_{g}/2}=1.4\cdot 10^{14}\, {\rm GeV}$.  In our  scenario, well below the inflaton mass scale  the low-energy effective field theory is given by the Standard Model (SM) that has the effective number of d.o.f. as $g_{*}=106.75$. It is reasonable to assume that all the SM particles originated from perturbative inflaton decay via the (Starobinsky) universal reheating mechanism, whose reheating temperature is known \cite{star1982,Vilenkin:1985md}:
\begin{equation}
\label{T1}
T_{reh}=\left( \fracmm{90}{\pi^{2}g_{*}}\right)^{1/4}\sqrt{\Gamma_{tot}M_{P}}=3\cdot 10^{9}\,{\rm GeV} \, . 
\end{equation}

On the other hand, the reheating temperature for heavy gravitino is given by \cite{Jeong:2012en}
\begin{equation}
\label{T2}
T_{reh}=1.5\cdot 10^{8}~{\rm GeV}\left(\fracmm{80}{g_{*}} \right)^{1/4}\left(\fracmm{m_{3/2}}{10^{12}\,{\rm GeV}} \right)^{3/2}\, . 
\end{equation}
Combining Eqs.~(\ref{T1}) and (\ref{T2}) we get the gravitino and Polonyi masses as follows: 
\begin{equation} \label{GM}
m_{3/2}=(7.7\pm 0.8)\cdot 10^{12}\, {\rm GeV} \quad {\rm and} 
\quad M_{A}= 2e^{-\VEV{J}}m_{3/2} > 2m_{3/2}~.
\end{equation}

\section{Primordial Black Holes in supergravity} 

\def\eV{\,{\rm eV}}
\def\keV{\,{\rm keV}}
\def\MeV{\,{\rm MeV}}
\def\GeV{\,{\rm GeV}}
\def\TeV{\,{\rm TeV}}
\def\sv{\left<\sigma v\right>}
\def\({\left(}
\def\){\right)}
\def\cm{{\,\rm cm}}
\def\K{{\,\rm K}}
\def\kpc{{\,\rm kpc}}
\def\beq{\begin{equation}}
\def\eeq{\end{equation}}
\def\bea{\begin{eqnarray}}
\def\eea{\end{eqnarray}}

PBHs may be formed in the early Universe by collapse of primordial density perturbations resulting from inflation, when these perturbations re-enter the horizon and are {\it large} enough, i.e. when gravity forces are larger than pressure, in general. Apart from being considered as another (non-particle) source for DM, some PBHs (of stellar mass type) are also considered as the candidates for the gravitational wave effects caused by the binary black hole mergers observed by LIGO/Virgo collaboration \cite{Abbott:2016blz,Abbott:2016nhf}.

The PBH {\it mass} $M_{PBH}$ is related to the perturbations scale $k$ by Carr's formula  \cite{Carr:1975qj}
\beq \label{carr}
M_{PBH}= \g \r \fracmm{4\p H^{-3}}{3}\approx M_{\odot}  \left( \fracmm{\g}{0.2}\right)
\left( \fracmm{g_*}{3.36}\right)^{-\frac{1}{6}}
\left( \fracmm{k/(2\p)}{3\cdot 10^{-9}{\rm Hz}}\right)^{-2}~~,
\eeq
whose coefficient $\g=3^{-3/2}\approx 0.2$, the (normalized) energy density is almost equal to the
(normalized) entropy density $g_*\approx 3.36$, and  $M_{\odot}$ stands for the Solar mass,
$M_{\odot}\approx 2\times 10^{33}$~g.

The PBHs {\it abundance} $f=\O_{PBH}/ \O_c$ is proportional to the {\it amplitude} of the scalar perturbations
$P_{\z}$, while for the LIGO events one finds $k/(2\p)\sim 10^{-9}~{\rm Hz}$, $P_{\z}\sim 10^{-2}$
and $f\sim 10^{-2}$, as the regards the orders of their magnitudes \cite{Abbott:2016blz,Abbott:2016nhf}. The
value of $10^{-9}~{\rm Hz}$ corresponds to $10^6~{\rm Mpc}^{-1}$.

In a single-field inflation, relevant perturbations are controlled by inflaton scalar potential, so that large fluctuations $P_{\cal R}\approx \fracmm{\k^2}{2\ve}\left(\fracmm{H}{2\p}\right)^2$ are produced when the slow roll parameter 
$\ve=r/16$ goes to zero, i.e. when the potential has  {\it a near-inflection point} where
\beq  \label{inflect}
V'\approx V'' \approx 0~.
\eeq

Since we want a copious PBH production along with observationally consistent CMB observables, we should
"decouple" these events, and demand the existence of another ("short") plateau in the scalar potential after the inflationary plateau towards the end of inflation. This is not the case for the Starobinsky inflation with the scalar potential (\ref{starp}), however, it can be easily achieved in a more general framework. Our supergravity framework in Sect.~3 is an example of such framework, because it leads to a single-field inflation governed by arbitrary function $J$, so that the associated inflaton scalar potential is given by $V=\frac{g^2}{2}(J')^2$. 

As an example, let us consider the inflaton scalar potential
\beq \label{exam}
\fracmm{V}{V_0}=\left( 1+\x - e^{-\a\f} -\x e^{-\b\f^2}\right)^2~~,  
\eeq
which is a deformation of the Starobinsky potential (\ref{starp}) with $\a=\sqrt{2/3}$ and the new real parameters $\b\geq 0$ and $\x\geq 0$. The Starobinsky potential (\ref{starp}) is recovered when $\x=0$. The scalar potential (\ref{exam}) falls into our supergravity framework, has Minkowski minimum at $\f=0$ and the inflationary plateau for large positive $\f$.  But, in addition, it also has  an inflection point in the "waterfall" region between the inflationary plateau and the Minkowski vacuum. Indeed, the conditions (\ref{inflect}) result in two equations, 
\beq\label{cond1}
\a e^{-\a\f} +2\x\b\f e^{-\b\f^2}=0
\eeq
and
\beq \label{cond2}
\a^2 e^{-\a\f} -2\x\b e^{-\b\f^2}+4\x\b^2 \f^2 e^{-\f\f^2}=0~~,
\eeq
respectively. They imply a quadratic equation on $\f$,
\beq 
\a\f +1 - 2\b \f^2=0~~,
\eeq	
whose solution is given by
\beq 
\f_*= \fracmm{\a + \sqrt{\a^2 +4\b}}{4\b}>0~~.
\eeq
Then the remaining condition above is solved by
\beq
\x = \fracmm{\a e^{-\a\f_* + \b\f_*^2}}{2\b\f_*}~~.
\eeq

Of course, there are many other possibilities to choose the scalar potential having the form of a real function squared. We just showed that it is possible to combine a viable (Starobinsky-like) inflation with a viable (stellar mass type) PBHs production in the context of supergravity.

\section{Conclusion}

Our results lead to the intriguing unifying picture of CDM, dark energy (positive cosmological constant)  and cosmological inflation, in which their parameter spaces are linked to each other.  This scenario also suggests the interesting phenomenology in the ultra high energy cosmic rays: super heavy Polonyi particles may  decay into the SM particles, as the secondaries, in top-bottom decays. Cosmological high energy neutrinos  from the primary and secondary decay channels can be tested by IceCube and ANTARES experiments. 

Another interesting outcome is that some  (stellar mass type) PBHs remnants produced from the supergravity fields can compose part of the CDM halo co-existing with gravitinos. In this scenario, gravitational wave signals from the PBHs mergers can be envisaged, with intriguing implications for LIGO/VIRGO experiments.  In short, gravitational wave experiments may provide us with precious indirect information about the scalar sector of the inflationary supergravity. 

Finally, the intriguing possibility exists for a unification of the inflaton in the vector multiplet, and the SUSY GUTs such as the flipped $SU(5)\times U(1)$ model arising from (Calabi-Yau) compactified heterotic superstrings or the intersecting D-branes.

\section*{Acknowledgements}
The work by SVK on gravity and supergravity is supported by the Competitiveness Enhancement Program of Tomsk Polytechnic University in Russia. This work is also supported by a Grant-in-Aid of the Japanese Society for Promotion of Science (JSPS) under 
No.~26400252, and the World Premier International Research Center Initiative (WPI Initiative), MEXT, Japan. SVK is grateful to the Institute for Theoretical Physics of Hannover University in Germany for kind hospitality extended to him during part of this investigation. The work by MK on physics of dark matter was supported by grant of Russian Science Foundation (project N-18-12-00213).



\begin{thebibliography}{99}

\bibitem{Ade:2015xua}
  P.~A.~R.~Ade {\it et al.} [Planck Collaboration],
  ``Planck 2015 results. XIII. Cosmological parameters,''
  Astron.\ Astrophys.\  {\bf 594}, A13 (2016).

\bibitem{Ade:2015lrj}
  P.~A.~R.~Ade {\it et al.} [Planck Collaboration],
  ``Planck 2015 results. XX. Constraints on inflation,''
  Astron.\ Astrophys.\  {\bf 594}, A20 (2016).
    
\bibitem{Array:2015xqh}
  P.~A.~R.~Ade {\it et al.} [BICEP2 and Keck Array Collaborations],
  ``Improved Constraints on Cosmology and Foregrounds from BICEP2 and Keck Array Cosmic Microwave Background 
  Data with Inclusion of 95 GHz Band,''
  Phys.\ Rev.\ Lett.\  {\bf 116},  031302 (2016).
 
\bibitem{Starobinsky:1980te}
  A.~A.~Starobinsky,  ``A new type of isotropic cosmological models without singularity,''
  Phys.\ Lett.\  {\bf 91B}, 99 (1980).

\bibitem{Ketov:2012yz}
S.~V. Ketov, ``Supergravity and Early Universe: the Meeting Point of Cosmology
  and High-Energy Physics,'' Int. J. Mod. Phys.
  {\bf A28}, 1330021 (2013).

 \bibitem{Ketov:2014qha}
 S.V.~Ketov and T.~Terada,
"Inflation in supergravity with a single chiral superfield",
Phys. Lett. {\bf B736}, 272 (2014).

 \bibitem{Ketov:2014hya}
  S.V.~Ketov and T.~Terada,
"Generic Scalar Potentials for Inflation in Supergravity with a Single Chiral Superfield"
JHEP {\bf 12}, 062 (2014).

  \bibitem{Farakos:2013cqa}
A.~Farakos, A.~Kehagias, and A.~Riotto,
"On the Starobinsky Model of Inflation from Supergravity",
Nucl. Phys. {\bf B876}, 187 (2013).

\bibitem{Ferrara:2013rsa}
  S.~Ferrara, R.~Kallosh, A.~Linde and M.~Porrati,
  ``Minimal Supergravity Models of Inflation,''
  Phys.\ Rev.\  {\bf D88},  085038 (2013).

\bibitem{Aldabergenov:2016dcu}
  Y.~Aldabergenov and S.~V.~Ketov,
  ``SUSY breaking after inflation in supergravity with inflaton in a massive vector supermultiplet,''
Phys. Lett. \ {\bf B761}, 115 (2016).
 
\bibitem{Aldabergenov:2017bjt} 
  Y.~Aldabergenov and S.~V.~Ketov,
  ``Higgs mechanism and cosmological constant in $N=1$
                        supergravity with inflaton in a vector multiplet,''
Eur. Phys. J. {\bf C77}, 233 (2017).

  \bibitem{Polonyi:1977pj}
J.~Polonyi,
``Generalization of the Massive Scalar Multiplet Coupling to the
  Supergravity'', Hungary Central Inst. Res. KFKI-77-93 preprint (1977, rec. July 1978), 5 pages, unpublished.

\bibitem{Khlopov:1984pf}
  M.~Y.~Khlopov and A.~D.~Linde,
  ``Is it easy to save the gravitino?,''
  Phys.\ Lett.\  {\bf 138B}, 265 (1984).
   
\bibitem{Khlopov:1993ye}
       M.~Y.~Khlopov, Y.~L.~Levitan, E.~V.~Sedelnikov and I.~M.~Sobol,
  ``Nonequilibrium cosmological nucleosynthesis of light elements: calculations by the Monte Carlo method,''
  Phys.\ Atom.\ Nucl.\  {\bf 57}, 1393 (1994) 
  [Yad.\ Fiz.\  {\bf 57}, 1466 (1994)].
 
 \bibitem{Kawasaki:2004qu}
 M.~Kawasaki, K.~Kohri and T.~Moroi,
  ``Big-Bang nucleosynthesis and hadronic decay of long-lived massive particles,''
  Phys.\ Rev.\  {\bf D71}, 083502 (2005).
 
\bibitem{Khlopov:2015oda}
  M.~Khlopov,
  ``Cosmological Probes for Supersymmetry,''
  Symmetry {\bf 7}, 815 (2015).
  
\bibitem{Banks:1993en}
  T.~Banks, D.~B.~Kaplan and A.~E.~Nelson,
  ``Cosmological implications of dynamical supersymmetry breaking,''
  Phys.\ Rev.\  {\bf D49}, 779 (1994).
 
\bibitem{deCarlos:1993wie}
  B.~de Carlos, J.~A.~Casas, F.~Quevedo and E.~Roulet,
  ``Model independent properties and cosmological implications of the dilaton and moduli sectors of 4-d strings,''
  Phys.\ Lett.\  {\bf B318}, 447 (1993).

\bibitem{Coughlan:1983ci}
  G.~D.~Coughlan, W.~Fischler, E.~W.~Kolb, S.~Raby and G.~G.~Ross,
  ``Cosmological problems for the Polonyi potential,''
  Phys.\ Lett.\  {\bf 131B}, 59 (1983).
 
\bibitem{Moroi:1994rs}
  T.~Moroi, M.~Yamaguchi and T.~Yanagida,
  ``On the solution to the Polonyi problem with ${\cal O}(10~{\rm TeV})$ gravitino mass in supergravity,''
  Phys.\ Lett.\  {\bf B342}, 105 (1995).

\bibitem{Kawasaki:1995cy}
  M.~Kawasaki, T.~Moroi and T.~Yanagida,
  ``Constraint on the reheating temperature from the decay of the Polonyi field,''
  Phys.\ Lett.\  {\bf B370}, 52 (1996).
  
\bibitem{Moroi:1999zb}
  T.~Moroi and L.~Randall,
  ``Wino cold dark matter from anomaly mediated SUSY breaking,''
  Nucl.\ Phys.\  {\bf B570}, 455 (2000).
 
\bibitem{Khlopov:1985jw}
  M.~Khlopov, B.~A.~Malomed and I.~B.~Zeldovich,
  ``Gravitational instability of scalar fields and formation of primordial black holes,''
  Mon.\ Not.\ Roy.\ Astron.\ Soc.\  {\bf 215}, 575 (1985).
  
\bibitem{Khlopov:2004tn}
  M.~Y.~Khlopov, A.~Barrau and J.~Grain,
  ``Gravitino production by primordial black hole evaporation and constraints on the inhomogeneity of the early universe,''
  Class.\ Quant.\ Grav.\  {\bf 23}, 1875 (2006).

\bibitem{Khlopov:2008qy}
  M.~Y.~Khlopov,
  ``Primordial black holes,''
  Res.\ Astron.\ Astrophys.\  {\bf 10}, 495 (2010).

\bibitem{miniPBH}
R.~V.~Konoplich, S.~G.~Rubin, A.~S.~Sakharov, M.~Yu.~Khlopov, 
"Formation of black holes in first-order phase transitions as a cosmological test of symmetry breaking mechanisms", 
Phys.~Atom.~Nucl. \ {\bf 62}, 1593 (1999).
 
\bibitem{miniPBH2} 
M.~Yu.~Khlopov, R.~V.~Konoplich, S.~G.~Rubin, A.~S.~Sakharov,
"First-order phase transitions as a source of black holes in the early universe",
Grav.~Cosmol. {\bf 6}, 153 (2000).

\bibitem{mf}  Y. Fujii and K-I. Maeda,
"The scalar-tensor theory of gravitation", Cambridge Univ. Press, Cambridge, 2007.

\bibitem{Galante:2014ifa}
 M.~Galante, R.~Kallosh, A.~Linde and D.~Roest,
"Unity of Cosmological Inflation Attractors",
Phys. Rev. Lett. {\bf 114},  141302 (2015).
  
  \bibitem{Nakada:2017uka}
  H.~Nakada and S.~V.~Ketov,
"Inflation from higher dimensions",
 Phys.\ Rev.\  {\bf D96}, 123530 (2017).

\bibitem{Ellis:2015pla} 
J.~Ellis, M.~A.~G. Garcia, D.~V. Nanopoulos and K.~A. Olive,
"Calculations of inflaton decays and reheating: with applications to no-scale inflation models",
 JCAP {\bf 1507}, 050 (2015).                    

\bibitem{Mukhanov:1981xt}
V.~F. Mukhanov and G.~V. Chibisov, 
"Quantum Fluctuations and a Nonsingular Universe",
JETP Lett. {\bf 33}, 532 (1981) [Pisma Zh. Eksp. Teor. Fiz. {\bf 33}, 549 (1981)].

\bibitem{Kaneda:2010ut}
S.~Kaneda, S.V. Ketov and N. Watanabe, 
"Fourth-order gravity as the inflationary model revisited",
Mod. Phys. Lett. {\bf A25}, 2753 (2010).

\bibitem{Wess:1992cp}
 J, Wess and J. Bagger,  "Supersymmetry and supergravity",
Princeton Univ. Press, Princeton, 1992.

\bibitem{Aldabergenov:2017hvp}
Y.~Aldabergenov and S.V.~Ketov,
"Removing instability of Polonyi-Starobinsky supergravity by adding FI term",
Mod. Phys. Lett. {\bf A33}, 1850032 (2018).

 \bibitem{Addazi:2017ulg} 
   A.~Addazi, S.~V.~Ketov and M.~Yu.~Khlopov,
  ``Gravitino and Polonyi production in supergravity,''
Eur. Phys. J. {\bf C78}, 642 (2017).

  \bibitem{star1982}  
 A.~A.~Starobinsky, 
 "Nonsingular model of the Universe with the quantum gravitational de Sitter stage and its observational consequences", 
 in the Proceedings of the 2nd International Seminar "Quantum Theory of Gravity" (Moscow, 13-15 October, 1981); 
 INR Press, Moscow 1982, p. 58 (reprinted in "Quantum Gravity", M.~A. Markov and P.~C. West Eds., Plemum Publ. Co., New York, 
 1984, p. 103).
  
\bibitem{Vilenkin:1985md}
  A.~Vilenkin, 
 "Classical and Quantum Cosmology of the Starobinsky Inflationary Model",
  Phys. Rev. {\bf D32}, 2511 (1985).


\bibitem{Jeong:2012en}
K.~S. Jeong and F.~Takahashi,
"A Gravitino-rich Universe",
JHEP {\bf 01}, 173 (2013).

\bibitem{Abbott:2016blz}
B.P.~Abbott {\it et al.}, [LIGO Scientific and Virgo Collaborations],
"Observation of Gravitational Waves from a Binary Black Hole Merger",
Phys. Rev. Lett. {\bf 116}, 061102 (2016).

\bibitem{Abbott:2016nhf}
B.P.~Abbott {\it et al.}, [LIGO Scientific and Virgo Collaborations],
"The Rate of Binary Black Hole Mergers Inferred from Advanced LIGO Observations Surrounding GW150914",
 arXiv:1602.03842 [astro-ph.HE].

\bibitem{Carr:1975qj}
B.J.~Carr, 
"The Primordial black hole mass spectrum",
Astrophys. J. {\bf 201}, 1(1975).


\end{thebibliography}
\end{document}